\documentclass[aps,prd,twocolumn,nofootinbib,amsmath,amssymb]{revtex4-1}

\usepackage{CJK}                     
\usepackage[dvips]{graphicx}         
\usepackage{mathptmx}                
\usepackage{bm}
\usepackage{dcolumn}                 
\usepackage{multirow}                
\usepackage{xcolor}

\usepackage{hyperref}
\hypersetup{,breaklinks=true,colorlinks=true,    
linkcolor=blue,citecolor=blue,filecolor=magenta,urlcolor=black}

\def\bea{\begin{eqnarray}} \def\eea{\end{eqnarray}}
\def\be{\begin{equation}} \def\ee{\end{equation}}
\def\bal#1\eal{\begin{align}#1\end{align}}
\def\pv{\bm{p}}

\def\eps{\varepsilon}
\def\la{\lambda}
\def\al{\alpha}
\def\om{\omega}
\def\ms{\,M_\odot}
\def\mmax{M_\text{max}}
\def\fm3{\,\text{fm}^{-3}}
\def\gc3{\,\text{g/cm}^3}
\def\mfm{\,\text{MeV}\,\text{fm}^{-3}}

\begin{document}

\title{The equation of state and radial oscillations of neutron stars}

\author{Ting-Ting Sun$^{1}$}
\author{Zi-Yue Zheng$^{1}$}
\author{Huan Chen$^{1}$}\email{Email:huanchen@cug.edu.cn}
\author{G. Fiorella Burgio$^{2}$}
\author{Hans-Josef Schulze$^{2}$}

\affiliation{
$^{1}$ School of Mathematics and Physics, China University of Geosciences,
Lumo Road 388, 430074 Wuhan, China
\\
$^{2}$ INFN Sezione di Catania and Dipartimento di Fisica,
Universit\'a di Catania, Via Santa Sofia 64, 95123 Catania, Italy
}

\begin{abstract}
We investigate radial oscillations of pure neutron stars and hybrid stars,
employing equations of state of nuclear matter
from Brueckner-Hartree-Fock theory,
and of quark matter from the Dyson-Schwinger quark model,
performing a Gibbs construction for the mixed phase in hybrid stars.
We calculate the eigenfrequencies and corresponding oscillation functions.
Our results for the zero points of the first-order radial oscillation frequencies
give the maximum mass of stable neutron stars,
consistent with the common criterion $dM/d\rho_c=0$.
Possible observations of the radial oscillation frequencies could help
to learn more about the equation of state,
predict the maximum mass of neutron stars more precisely,
and indicate the presence of quark matter.
\end{abstract}



\maketitle

\section{Introduction}

Neutron stars (NSs),
the densest observable stars in the Universe,
are natural laboratories for the study of cold dense nuclear matter.
Theoretically, the equation of state (EOS) of nuclear matter is the key input
which determines the structure and properties of NSs.
Unfortunately, due to the nonperturbative nature of the strong interaction,
our knowledge about the EOS of dense nuclear matter is still insufficient,
especially at densities much higher than the nuclear saturation density,
where deconfined quark matter may probably be present \cite{Weber2005}.

Many efforts have been made to constrain the EOS from the observation of NSs.
For the static and spherical case,
one can obtain the equilibrium structure of NSs by solving the
Tolman-Oppenheimer-Volkov (TOV) equations combined with the EOS,
thus predicting mass-radius-central density relations of NSs.
The most recent observations have been performed by the NICER
(Neutron Star Interior Composition Explorer) mission,
which reported a Bayesian parameter estimation of the mass and equatorial radius
of the millisecond pulsar PSR J0030+0451 \cite{nicer3,nicer4}.
Additional constraints are imposed by the largest mass observed up to now,
$2.14^{+0.10}_{-0.09}\,\ms$ for the object PSR J0740+6620 \cite{cromartie},
and by recent analyses of the NS merger event GW170817,
which indicate an upper limit on the maximum mass of about $2.2-2.3\,\ms$
\cite{shiba17,marga17,rezz18,Shibata19}.

Theoretically, the difference of mass-radius relations between hybrid stars (HSs)
and pure NSs is not significant,
and depends sensitively on the various models adopted for describing
nuclear and quark matter \cite{Alford2009apj}.
Therefore, one cannot yet disentangle HSs from pure NSs
on the basis of the current observations.

NSs also undergo different kinds of mechanical deformations,
e.g., radial and non-radial oscillations, glitches,
and even those resulting from NS mergers
\cite{TheLIGOScientific2017,Monitor2017}.
These produce many kinds of electromagnetic and gravitational-wave (GW) signals,
and also indicate the internal structure of NSs.
In this work we will mainly concentrate on the NS radial oscillations,
which were first studied
in general relativity by Chandrasekhar \cite{Chandrasekhar1964}.
Subsequently, many investigations were carried out
\cite{Chanmugam1977Radial,Glass1983,Chanmugam1992,Gondek1997,Sahu2001,
Kokkotas2000,Gupta2002,Brillante2014,Panotopoulos2017,Flores2017,Sagun2020}.
Altough radial oscillations cannot lead to direct emission of GW radiation,
they can couple with and amplify GWs \cite{Passamonti2005,Passamonti2007},
and therefore could be observed in GW signals.
They can also modulate the short gamma ray burst (SGRB) from the hypermassive NSs
formed after the merger of two NSs,
and the frequency could thus be observed in SGRB \cite{Chirenti2019}.
In this work, we will investigate how the frequencies of the radial oscillations
depend on the internal structure
and composition of the emitting source, thus identifying  pure NSs and HSs.

Many theoretical tools and models have been proposed to study the EOS of NSs,
see, e.g., \cite{2018Chap6} for a review.
For nuclear matter in the hadron phase,
popular EOSs are based on
relativistic mean field models \cite{Ring1996},
phenomenological models based on energy-density functional theory
with generalized Skyrme effective forces
\cite{BSK},
Brueckner-Hartree-Fock (BHF) theory
\cite{Li2012,HeYeTongLang2013,Fukukawa2015,Lu19},
the variational method (APR) \cite{apr1998},
the self-consistent Green's functions approach
\cite{Carbone2013},
and chiral effective field theory 
\cite{Hebeler2010,Coraggio2014,Wellenhofer2014,Drischler2015}.
For quark matter, EOSs are mainly obtained with
the MIT bag model \cite{Chodos1974},
the Nambu-Jona-Lasinio (NJL) model \cite{Buballa2005,Klahn2013,Klahn2015},
the perturbative QCD \cite{Kurkela2009,Fraga2014,Jimenez2019},
and the Dyson-Schwinger equations (DSEs)
\cite{Roberts1994,Alkofer2000,Chen2011,Chen2012,Chen2015}.

In this work, we model nuclear matter with the BHF theory,
which is based on realistic two- and three-body forces that describe accurately
nucleon scattering data in free space and the properties of the deuteron.
Moreover the BHF approach is able to describe properly the properties
of symmetric nuclear matter at the saturation density
\cite{Li2012,HeYeTongLang2013,Fukukawa2015,Wei20}.
For quark matter, we adopt the Dyson-Schwinger quark model
\cite{Chen2011,Chen2015}.
The phase transition between the confined and
deconfined phase is modelled with the Gibbs condition
\cite{Glendenning1992,Chen2011}.
In this framework,
the maximum masses of the pure NSs and HSs fulfill the two-solar-mass constraint
\cite{demorest2010,heavy2,fonseca16,cromartie}.

The work is organized as follows.
In Sec.~\ref{s:eos} we briefly describe the formalism for the EOSs, i.e.,
the BHF theory for the hadron phase and the DSEs for the quark phase.
In Sec.~\ref{s:osc} we introduce the TOV and the Sturm-Liouville
eigenvalue equations for the internal structure and radial oscillations of NSs.
Numerical results are given in Sec.~\ref{s:res},
and we draw the conclusions in Sec.~\ref{s:end}.
We use natural units $c=\hbar=1$ throughout.

\section{Equation of state}
\label{s:eos}

\subsection{Nuclear matter}

In the BHF theory,
the key element to describe the dense nuclear matter is the $G$-matrix,
which satisfies the Bethe-Goldstone equation \cite{Baldo:1999fbt}
\be\label{eq.g}
 G[E;\rho] = V + \sum_{k_a,k_b>k_F} V
 \frac{| k_a,k_b \rangle Q \langle k_a,k_b |}{E-e(k_a)-e(k_b)} G[E;\rho] \:,
\ee
where $E$ is the starting energy,
$\rho$ is the nucleon number density,
$V$ is the interaction potential,
and $Q$ is the Pauli operator.
The single-particle (s.p.) energy of the nucleon is
\be\label{eq.e}
  e(k) = e(k;\rho) = \frac{k^2}{2m} + U(k,\rho) \:,
\ee
where
\be\label{eq.u}
 U(k;\rho) = \sum_{k'\leq k_F}
 \langle kk' | G[e(k)+e(k');\rho] | kk'\rangle_A
\ee
is the s.p.~potential under the continuous choice.

By solving the equations (\ref{eq.g},\ref{eq.e},\ref{eq.u}),
one can obtain the $G$-matrix and then the energy per nucleon of nuclear matter
\be\label{energy}
 \frac{B}{A} = \frac{3}{5} \frac{k_F^2}{2m} +
 \frac{1}{2\rho} \sum_{k,k'<k_F}
 \langle kk' | G[e(k)+e(k');\rho] | kk' \rangle_A \:.
\ee
In this work we use the Bonn~B (BOB)
\cite{Machleidt:1987hj,Machleidt:1989tm}
and Argonne $V_{18}$ (V18) \cite{Wiringa:1994wb}
nucleon-nucleon potentials,
supplemented with compatible microscopic three-body forces \cite{zuo2002,Li2008}.
This is a common prescription adopted in the BHF approach,
and allows to reproduce correctly
the saturation point of symmetric nuclear matter \cite{Baldo:1999fbt}.

In order to study the structure of the NS core,
we have to calculate the composition and the EOS of cold,
neutrino-free, catalyzed matter,
imposing that it contains charge-neutral matter consisting of
neutrons, protons, and leptons ($e^-$, $\mu^-$) in beta-equilibrium.
The output of the many-body calculation is the energy density
of lepton/baryon matter as a function of the different partial densities
$\rho_i$ of the species $i=n,p,e,\mu$,
\bal
 \eps(\rho_n,\rho_p,\rho_e,\rho_\mu) =&
 (\rho_nm_n+\rho_pm_p) + (\rho_n+\rho_p)\frac{B}{A}(\rho_n,\rho_p)
\nonumber \\
  & + \eps_e(\rho_e) + \eps_\mu(\rho_\mu) \:,
\label{eq:ea}
\eal
where $m_i$ are the corresponding masses,
$B/A(\rho_n,\rho_p)$ is the energy per nucleon of asymmetric nuclear matter,
and $\eps_e$ and $\eps_\mu$ are the energy densities of electrons and muons,
which are usually considered as noninteracting
(we use ultrarelativistic and relativistic expressions
for the energy densities of electrons $\eps(\rho_e)$ and muons $\eps(\rho_\mu)$,
respectively \cite{shapiro}).

Given the large computational efforts of the microscopic calculations,
we have used the parabolic approximation \cite{hypns2,bom3}
of the energy per particle of asymmetric nuclear matter in Eq.~(\ref{eq:ea}),
with the symmetry energy calculated simply as the difference
between the energy per particle of pure neutron matter
and symmetric nuclear matter,
\be
 E_\text{sym}(\rho) \approx
 E(\rho_n=\rho,\rho_p=0) - E(\rho_n=\rho/2,\rho_p=\rho/2) \:.
\label{e:sym}
\ee

Once the energy density, Eq.~(\ref{eq:ea}), is known,
the various chemical potentials can be computed,
\be
 \mu_i = {\partial \eps \over \partial \rho_i} \:,
\ee
and solving the equations for beta-equilibrium,
\bal
& \mu_p+\mu_e=\mu_n=\mu_B \:,
\\
& \mu_e=\mu_\mu=\mu_C     \:,
\eal
being $\mu_B$ the baryon number chemical potential
and $\mu_C$ the electric charge chemical potential,
corresponding to the only two conserved charges,
along with the charge neutrality,
\be
 \rho_p = \rho_e + \rho_\mu  \:,
\ee
allows one to find the equilibrium composition $\rho_{i}$
at fixed baryon density $\rho$, and finally the EOS,
\be
 p(\eps) = \rho^2 {d\over d\rho}
 {\eps(\rho_i(\rho))\over d\rho}
 = \rho {d\eps \over d\rho} - \eps
 = \rho \mu_n - \eps \:.
\ee

We notice that the above mentioned theoretical methods provide EOSs
for homogeneous nuclear matter,
$\rho > \rho_t \approx 0.08\,\text{fm}^{-3}$.
For the low-density inhomogeneous part
we adopt the well-known Negele-Vautherin EOS \cite{Negele1971}
for the inner crust in the medium-density regime
($0.001\,\text{fm}^{-3} < \rho < \rho_t$),
and the ones by Baym-Pethick-Sutherland \cite{Baym1971}
and Feynman-Metropolis-Teller \cite{Feynman1949} for the outer crust
($\rho < 0.001\,\text{fm}^{-3}$).

However, the BHF approach is a non-relativistic theory, and
the above EOS predicts a superluminal speed of sound $v_s^2=dp/d\eps>c^2$
at a few times of the saturation density \cite{Luo2019},
close to the central density of the most massive NSs.
As a simple remedy,
we truncate the EOS at $v_s^2=c^2$
and keep the speed of sound constant at higher densities.
Later we will investigate the effects of such a modification
on the structure and radial oscillations of NSs.

\subsection{Quark matter}

As discussed above,
it is reasonable to assume that a phase transition from hadronic matter
to QM occurs in NSs.
Quarks are usually described in relativistic theories,
and most models predict that $v_s^2<1/3$.
Therefore the speed of sound drops with the phase transition to QM,
and thus the causality requirement is easily fulfilled
if the phase transition occurs before the speed of sound approaches unity
in hadronic matter,
which could also constrain parameters in the models.

As in Ref.~\cite{Luo2019},
we use the Dyson-Schwinger model (DSM) \cite{Chen2011} to describe QM.
In the DSM, one starts from
the quark propagator at finite chemical potentials $S(p;\mu)$,
which satisfies the Dyson-Schwinger equation
\bal\label{e:dse}
 & S(p;\mu)^{-1} = Z_2 (i\gamma\cdot \tilde{p}+m_q)
\\ &
 + Z_1 g^2(\mu) \int \frac{d^4q}{(2\pi)^4}
D_{\rho\sigma}(k;\mu) \frac{\lambda^a}{2} \gamma_\rho
 S(q;\mu) \frac{\lambda^a}{2} \Gamma_\sigma(q,p;\mu) \:,
\nonumber
\eal
where
$\tilde p \equiv (\pv,p_4+i\mu)$,
$D_{\rho\sigma}(k\equiv p-q;\mu)$
is the full gluon propagator,
$\Gamma_\sigma(q,p;\mu)$ is the effective quark-gluon vertex, and
$Z_1$ and $Z_2$ are the renormalization constants for the quark-gluon vertex
and the quark wavefunction.
With a given ansatz for the quark-gluon vertex and gluon propagator,
one can solve the equation and obtain the quark propagator.
In Refs.~\cite{Chen2011,Luo2019},
the so-called rainbow approximation and a
chemical-potential-modified Gaussian-type effective interaction were used,
see Ref.~\cite{Chen2011} for details.

The EOS for cold QM is given by the $q=u,d,s$ quark propagator
at zero temperature as in Refs.~\cite{Chen2008,Klahn2009},
\bal\label{dsrho}
 \rho_q(\mu_q) &= 6\int \frac{d^4p}{(2\pi)^4}
 \mathop{\text{tr}_D}[-\gamma_4S_q(p;\mu_q)] \:,
\\
  p_q(\mu_q) &= p_q(\mu_{q,0}) + \int_{\mu_{q,0}}^{\mu_q} d\mu \rho_q(\mu) \:.
\label{dsp}
\eal
The total density and pressure of the QM are given
by summing contributions from all flavors,
and the pressure of QM at zero density is taken
as a phenomenological bag constant \cite{Wei2017},
\be\label{BDS}
 B_\text{DS} = -\sum_{q=u,d,s} p_q(\mu_{q,0}) \:,
\end{equation}
which is set to $90\mfm$ \cite{Chen2012,Chen2015,Wei2017}.

In the pure quark phase,
the beta-equilibrium and electrical neutrality are expressed as
\bal
 & \mu_d = \mu_u + \mu_e = \mu_u + \mu_\mu = \mu_s \:,
\\
 & \frac{2n_u-n_d-n_s}{3} - \rho_e - \rho_\mu = 0 \:.
\eal
The phase transition from nuclear phase to quark phase
is considered as the Gibbs construction \cite{Glendenning1992,Chen2011},
where the chemical and mechanical equilibrium between both phases
are expressed as
\bal\label{chemical_equilibrium}
 & \mu_n = \mu_p+\mu_e = \mu_u+2\mu_d  \:,
\\
 & p_H(\mu_e,\mu_n) = p_Q(\mu_e,\mu_n) = p_M(\mu_n) \:.
\eal
In the mixed phase,
the hadron phase and quark phase are electrically charged separately,
while it remains globally neutral,
\be\label{chi}
 \chi\rho_Q + (1+\chi)\rho_H = 0 \:,
\ee
where $\chi$ is the volume fraction of QM in the mixed phase.
Consequently,
the baryon number density $\rho_M$ and energy density $\eps_M$
of the mixed phase are
\bal\label{energy_density}
 \rho_M &= \chi\rho_Q + (1+\chi)\rho_H \:,
 \\
 \eps_M &= \chi\eps_Q + (1+\chi)\eps_H \:.
\eal
With the above phase transition,
the EOS is continuous at the phase transition onset,
different from the Maxwell phase transition considered
in Ref.~\cite{Pereira2018}.
However, the corresponding speed of sound drops discontinuously
at the phase-transition point \cite{Luo2019}.

\section{Hydrostatic equilibrium structure and radial oscillations}
\label{s:osc}

Due to the strong gravitational field in NSs,
their structure and dynamical evolution are ruled
by the Einstein equation in general relativity,
\be\label{field}
 R_{\mu\nu} - \frac{1}{2}g_{\mu\nu}R = 8\pi G T_{\mu\nu} \:,
\ee
where $R_{\mu\nu}$ is the Ricci tensor,
$R$ is the Ricci scalar, and
$G$ is the gravitational constant.
The energy-momentum tensor is
\be\label{tensor}
 T_{\mu\nu} = pg_{\mu\nu} + (p+\eps)u_\mu u_\nu \:,
\ee
where $g_{\mu\nu}$ is the metric tensor,
$p$ is the pressure,
$\eps$ is the energy density,
and $u_\mu$ is the four-velocity.
For simplicity, we consider static spherically symmetric stars,
described by the Schwarzschild metric
\cite{Chandrasekhar1964}
\be\label{e:ds2}
 ds^2 = e^{\nu(r)}dt^2 - e^{\lambda(r)}dr^2 -
 r^2(d\theta^2+\sin^2\!\theta d\varphi^2) \:,
\ee
where $e^{\nu(r)}$ and $e^{\lambda(r})$ are metric functions.
By solving the Einstein field equation with the above metric,
one obtains the TOV equations
\cite{Oppenheimer:1939ne,Tolman:1939jz}
for the equilibrium structure of NSs,
\bal\label{dpdr}
 \frac{dp}{dr} &= -G\frac{(\eps+p)(m+4\pi r^3p)}{r^2(1-2Gm/{r})} \:,
\\
 \frac{dm}{dr} &= 4\pi r^2\eps \:,
\eal
and correspondingly the metric functions
\bal\label{A}
 e^{\la(r)} &= (1-2Gm/r)^{-1} \:,
\\
 {\nu(r)} &=
 -2G \int_r^\infty dr' \frac{e^{\la(r')}}{r'^2}
 \left( m + 4\pi r'^3 p \right) \:.
\eal
Combining with the EOS $p(\eps)$ of the matter,
one can solve the TOV equations
for the initial conditions $m(r=0)=0$ and $p(r=0)=p_c$,
where $p_c$ is the central pressure.
The surface radius is defined by $p(R)=0$
and the corresponding NS mass is $M=m(R)$.

\begin{figure}[t]
\centerline{\includegraphics[width=0.5\textwidth]{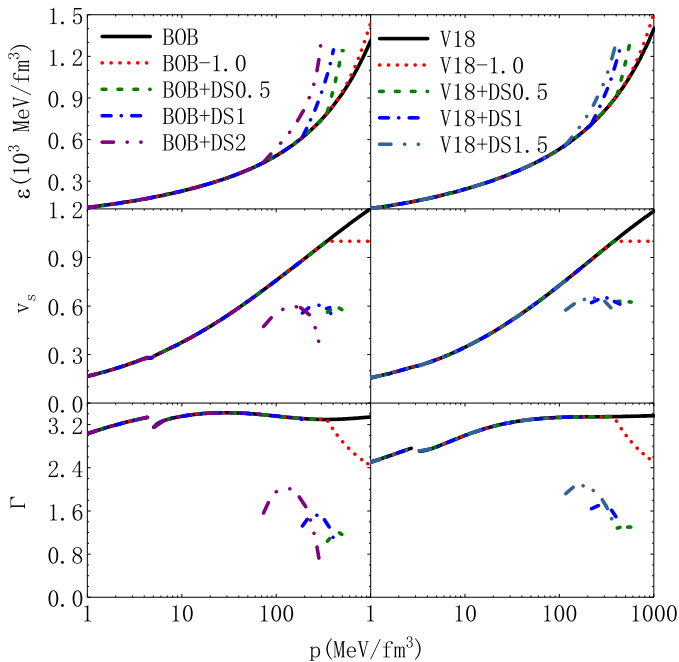}}
\vskip-3mm
\caption{
The energy density (upper panels),
speed of sound (central panels),
and adiabatic index (lower panels) of NS matter
as functions of pressure with different EOSs.
See the text for a detailed description of the notation.
}
\label{f:sound}
\end{figure}

\begin{figure}[t]
\centerline{\includegraphics[width=0.44\textwidth]{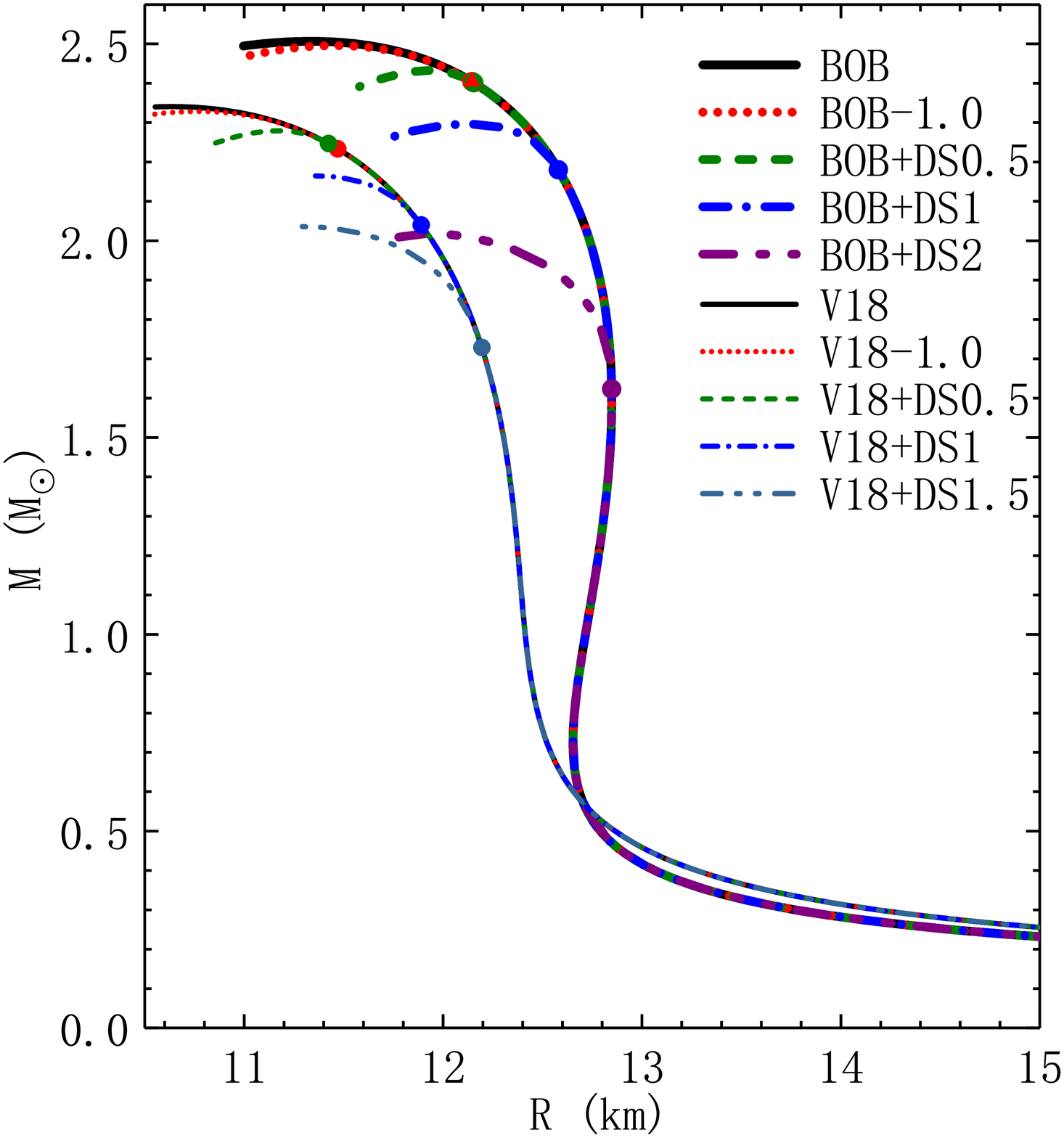}}
\vskip-2mm
\caption{
The mass-radius relations of NSs obtained with different EOSs.
The markers represent the minimum HS mass
(QM onset)
for each EOS.
}
\label{f:mr}
\end{figure}

Also the radial oscillation properties can be obtained from the
Einstein field equation \cite{Harrison:1965,Bardeen1966A}
based on the static equilibrium structure.
Consider a spherically symmetric system with only radial motion,
where the metric Eq.~(\ref{e:ds2}) is now time dependent.
Small perturbations are described by
$\xi\equiv\Delta r/r$,
where $\Delta r$ is the radial displacement,
and the corresponding Lagrangian perturbation of the pressure
$\eta\equiv\Delta p/p$
\cite{Chanmugam1977Radial,Chanmugam1992,Sagun2020}.
The time dependence of all the perturbations can be obtained
as a summation of eigenmodes
$\xi_i$ and $\eta_i \varpropto e^{i\om_i t}$
that are solutions of the system of differential equations
\cite{Chanmugam1977Radial,Gondek1997},
\bal
 \frac{d\xi}{dr} = & -\frac{1}{r}
 \Big( 3\xi + \frac{\eta}{\Gamma} \Big)
 -\frac{dp}{dr} \frac{\xi}{p+\eps} \:,
\label{e:xi}
\\
 \frac{d\eta}{dr} = &
 \frac{\xi}{p} \bigg[ \om^2 e^{\lambda -\nu} (p+\eps)r - 4\frac{dp}{dr}
\nonumber\\
 &\hskip5mm + \Big(\frac{dp}{dr}\Big)^2 \frac{r}{p+\eps}
 -8\pi G e^\lambda(p+\eps)pr \bigg]
\nonumber\\
 & + \eta \bigg[ \frac{dp}{dr}\frac{1}{p+\eps}
 -4\pi G e^\lambda(p+\eps)r \bigg] \:,
\label{e:dp}
\eal
where $\om=2\pi f$ is the eigenfrequency of radial oscillation,
and
\be\label{e:gamma}
 \Gamma = \left(1+\frac{\eps}{p}\right) v_s^2
\ee
is the adiabatic index.
Two boundary conditions are required in addition \cite{Gondek1997}.
The condition at the center is
\be\label{e:r0}
 \eta(0) = -3\Gamma\xi \:,
\ee
while the perturbation of the pressure should vanish at the surface,
\be\label{e:rR}
 \eta(R) = 0 \:.
\ee

Eqs.~(\ref{e:xi})
to (\ref{e:rR})
are the Sturm-Liouville eigenvalue equations for $\om$.
The solutions provide the discrete eigenvalues $\om_i^2$.
For a given NS, they can be ordered as
$\om_1^2<\om_2^2<\ldots<\om_n^2$,
where $n$ is the number of nodes.
Negative $\om^2$ indicate unstable oscillations
and thus $\om^2=0$ is the critical condition for the stability of NSs
under radial perturbations.

\begin{table*}[t]
\caption{
Characteristic static properties of neutron stars and HSs
with different masses
$\mmax [\ms]$, $2.0\ms$, and $1.4\ms$:
Central nucleon number density $\rho_c\;[\fm3]$ and pressure $p_c\;[\mfm]$,
radius $R\;[\text{km}]$,
and compactness parameter $\beta={GM}/{R}$.
$M_\text{min}\;[\ms]$ is the minimum NS mass of a given (hybrid) EOS.
Note that only the DS1.5 and DS2 QM EOSs
allow hybrid $2\ms$ configurations.
No hybrid $1.4\ms$ stars exist.
}
\def\myt#1{\multicolumn{1}{c|}{#1}}
\begin{ruledtabular}
\begin{tabular}{l|c|ccccc|cccc|cccc}
\myt{$M$}& $M_\text{min}$
         & \multicolumn{5}{c|}{$\mmax$}
         & \multicolumn{4}{c|}{$2.0\ms$}
         & \multicolumn{4}{c}{$1.4\ms$} \\
\hline
EOS      &     & $\mmax$
         & $\rho_c$ & $p_c$ & $R$ & $ \beta$
         & $\rho_c$ & $p_c$ & $R$ & $ \beta$
         & $\rho_c$ & $p_c$ & $R$ & $ \beta$ \\
\hline
BOB      &     & 2.51& 0.884& 819.4& 11.34& 0.326& 0.510& 132.7& 12.74& 0.232& 0.384& 51.0& 12.83& 0.161\\
BOB-1.0  & 2.40 & 2.50& 0.881& 726.2& 11.46& 0.321&      &      &      &      &      &     &      &      \\
BOB+DS0.5& 2.40 & 2.43& 0.860& 449.2& 11.93& 0.301&      &      &      &      &      &     &      &      \\
BOB+DS1  & 2.18 & 2.30& 0.838& 333.9& 12.12& 0.279&      &      &      &      &      &     &      &      \\
BOB+DS2  & 1.62 & 2.02& 0.910& 268.3& 11.94& 0.249& 0.750& 208.9& 12.21& 0.242&      &     &      &      \\
\hline
V18      &     & 2.34& 1.010& 960.9& 10.63& 0.345& 0.632& 199.6& 11.94& 0.247& 0.452& 65.3& 12.34& 0.167\\
V18-1.0  & 2.23 & 2.33& 1.002& 830.9& 10.77& 0.319&      &      &      &      &      &     &      &      \\
V18+DS0.5& 2.25 & 2.28& 0.983& 547.7& 11.17& 0.301&      &      &      &      &      &     &      &      \\
V18+DS1.0& 2.04 & 2.16& 0.966& 417.2& 11.36& 0.281&      &      &      &      &      &     &      &      \\
V18+DS1.5& 1.73 & 2.04& 1.001& 374.9& 11.29& 0.266& 0.785& 249.4& 11.71& 0.252&      &     &      &
\end{tabular}
\end{ruledtabular}
\label{t:res}
\end{table*}

\section{Numerical results}
\label{s:res}

\begin{figure}[t]
\vskip-1mm
\centerline{\includegraphics[width=0.51\textwidth]{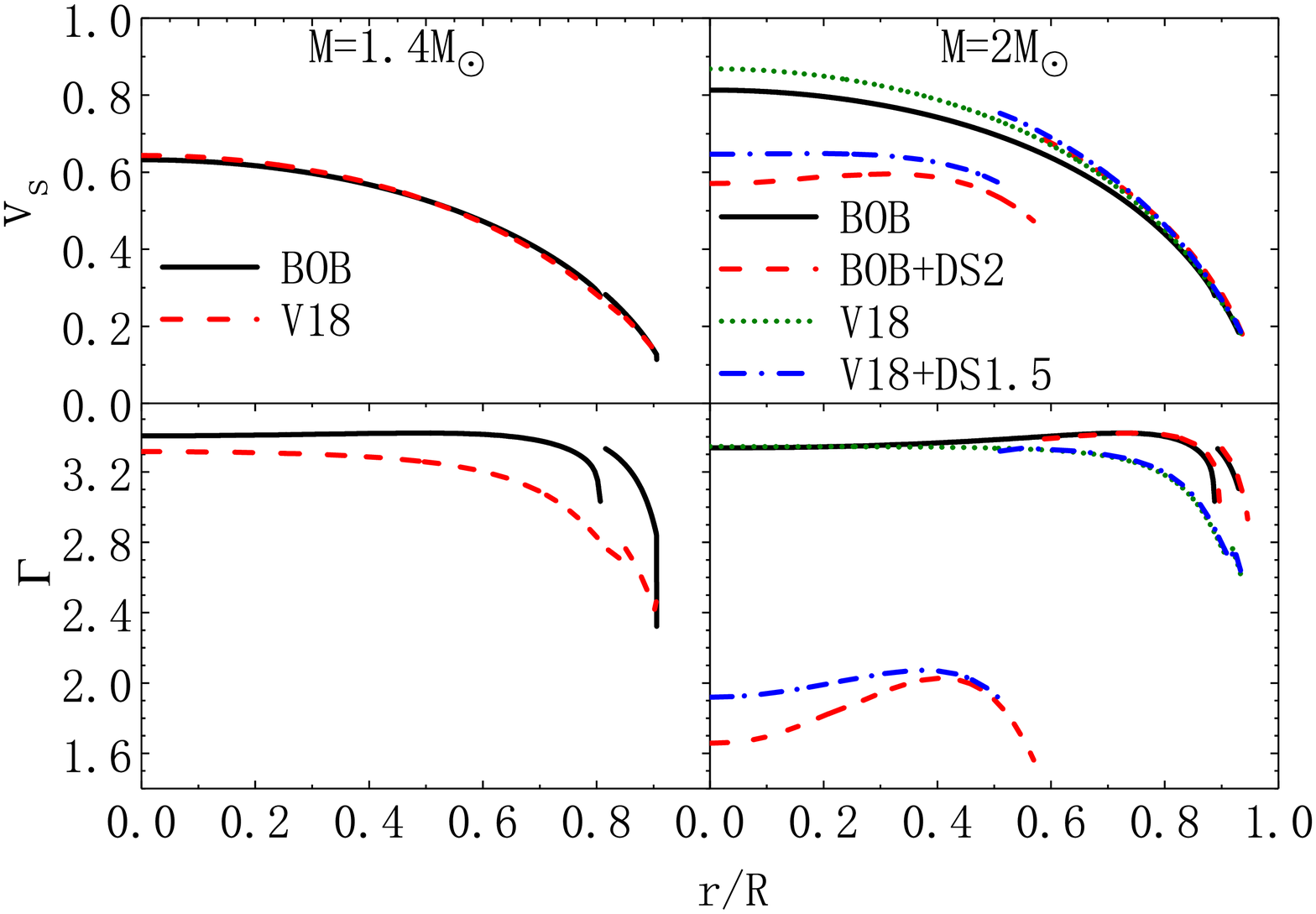}}
\vskip-5mm
\caption{
Speed of sound (upper panels)
and adiabatic index $\Gamma$ (lower panels)
in NSs with $1.4\ms$ (left panels)
and $2.0\ms$ (right panels),
for various EOSs.
}
\label{f:gamma}
\end{figure}

\begin{figure}[t]
\vskip-0mm
\centerline{\includegraphics[width=0.51\textwidth]{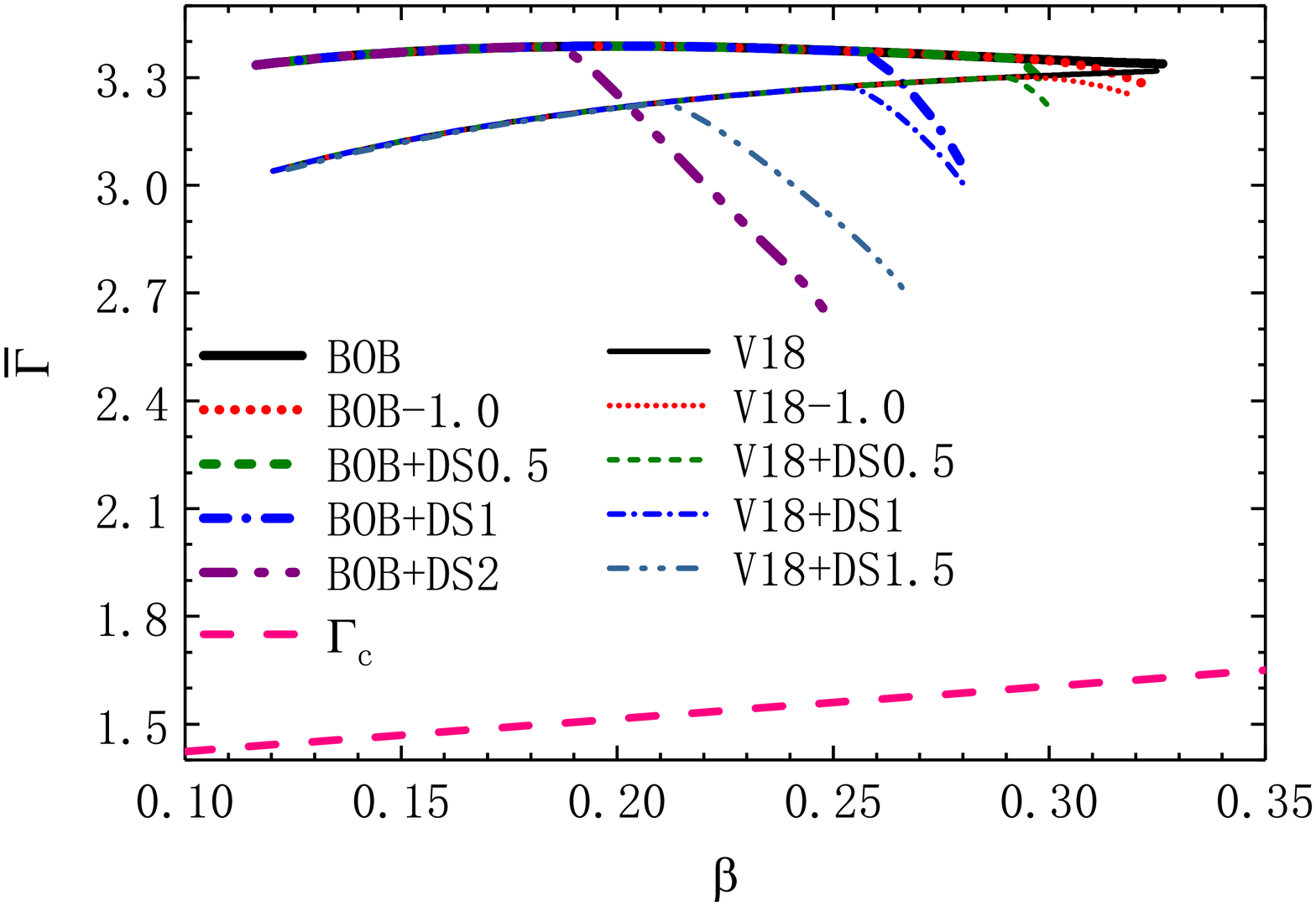}}
\vskip-4mm
\caption{
The dependence of the averaged adiabatic index $\bar\Gamma$
on the compactness parameter $\beta$ with various EOSs,
in comparison with the critical value Eq.~(\ref{e:gc})
(dashed curve).}
\label{f:gamma-beta}
\end{figure}

\subsection{Equilibrium structure of neutron stars}

As stated above, we consider two kinds of EOSs,
corresponding to pure NSs and HSs respectively.
Details can be found in Refs.~\cite{Li2008,Chen2011}.

For the pure NS EOS we use the BOB and V18 BHF EOSs discussed before.
The energy density, speed of sound,
and adiabatic index as functions of the pressure
are shown in Fig.~\ref{f:sound}.
At large pressure the speed of sound exceeds the speed of light
and violates causality.
As a simple remedy,
we truncate the EOS at $v_s/c=1$ and keep it constant at higher densities.
The modified results shown in the figure are labeled as
``BOB-1.0" or ``V18-1.0".
One can see that the EOSs are softened with the reduction of the speed of sound.

For the EOS in HSs,
labeled as V18/BOB+DS$\al$,
we combine the EOS of nuclear matter
and the DSM EOS of QM
with different parameters $\al=0.5,1,1.5,2$,
representing the strength of the in-medium modification of the
Gaussian-type effective interaction,
see Refs.~\cite{Chen2011,Luo2019} for details.
The EOS, the speed of sound,
and the adiabatic index are also shown in Fig.~\ref{f:sound}.
At the onset of the mixed phase,
there is a discontinuous decrease of the speed of sound and the adiabatic index
due to the emergence of new degrees of freedom,
similar to the onset of muons at low density.
In the mixed phase, the speed of sound is thus much lower,
without causality violation,
and depends strongly and non-monotonously on the pressure.
The EOS is also much softened by the phase transition.

The corresponding mass-radius relations of NSs are shown in Fig.~\ref{f:mr},
obtained in the standard way by solving the TOV equations for betastable and
charge-neutral matter.
We remark that the effect of flattening the V18/BOB EOS at $v_s/c=1$
is very small and the value of the maximum mass $\mmax=2.34/2.51\ms$
is larger than the current observational lower limits
\cite{demorest2010,heavy2,fonseca16,cromartie}.
Regarding the radius,
we found in \cite{drago4,Wei20} that for the V18/BOB EOS
the value of a 1.4-solar-mass NS is $R_{1.4}=12.3/12.8\;$km,
which fulfills the constraints derived from
the tidal deformability in the GW170817 merger event,
$R_{1.36}=11.9\pm1.4\;$km \cite{merger2},
see also similar compatible constraints on masses and radii
derived in Refs.~\cite{%
marga17,  
Ruiz18,      
Most18,      
rezz18,  
Shibata19,   
Most20,      
Shao20}.     
The V18/BOB EOSs are also compatible with estimates of the mass and radius
of the isolated pulsar PSR J0030+0451
observed recently by NICER,
$M=1.44^{+0.15}_{-0.14}\ms$ and $R=13.02^{+1.24}_{-1.06}\,$km
\cite{nicer3}, or
$M=1.36^{+0.15}_{-0.16}\ms$ and $R=12.71^{+1.14}_{-1.19}\,$km
\cite{nicer4}.

The phase transition
leads to smaller values of the maximum mass on the mass-radius curves,
obtained with the commonly used stability criterion
$dM/d\rho_c=0$ \cite{Harrison:1965,shapiro}.
In Table~\ref{t:res} we also list some characteristic static properties of NSs
with different EOSs.
One notes that for a too early onset of the quark phase the maximum mass
becomes too low.
The condition $\mmax>2.1\ms$ yields for example the constraint
$\al< 1.7/1.3$ for the BOB/V18+DS$\al$ EOS,
see also Refs.~\cite{Chen2011,Wei19}.

In Fig.~\ref{f:gamma} we show the profiles of the speed of sound
and the adiabatic index in NSs
with two different masses $M/\ms=1.4,2.0$.
In both cases the hadronic EOS never becomes superluminal.
One can see that due to the phase transition,
the speed of sound and adiabatic index in the inner core of heavy NSs
are reduced strongly,
and even decrease as approaching the center.
There are also the discontinuities
due to the emergence of the mixed phase in HSs,
similar as at the onset of muons in the outer layer
but quantitatively much larger.

It was often discussed in the literature \cite{Moustakidis2017}
that the stability of NSs depends crucially
on the averaged adiabatic index
\be\label{e:ave}
 \bar\Gamma \equiv
 \frac{\int_0^R dr r^2 e^{(\lambda+3\nu)/2}p \,\Gamma}
      {\int_0^R dr r^2 e^{(\lambda+3\nu)/2}p} \:.
\ee
In Ref.~\cite{Chandrasekhar1964}
Chandrasekhar gave a critical value as stability criterion,
\be\label{e:gc}
 \bar\Gamma_c = \frac{4}{3} + \frac{38}{42}\beta \:.
\ee
In Fig.~\ref{f:gamma-beta} we compare our results of the averaged adiabatic
index with various EOSs and with this criterion.
One finds that with emergence of the mixed phase in HSs,
the averaged adiabatic index decreases with the mass, compared to pure NSs.
However, in all cases it remains much larger than the critical value.
Therefore, Eq.~(\ref{e:gc}) could be regarded
as a necessary condition for the stability of NSs,
but far from the values obtained by realistic EOSs.

\begin{figure}[t]
\centerline{\includegraphics[scale=0.33]{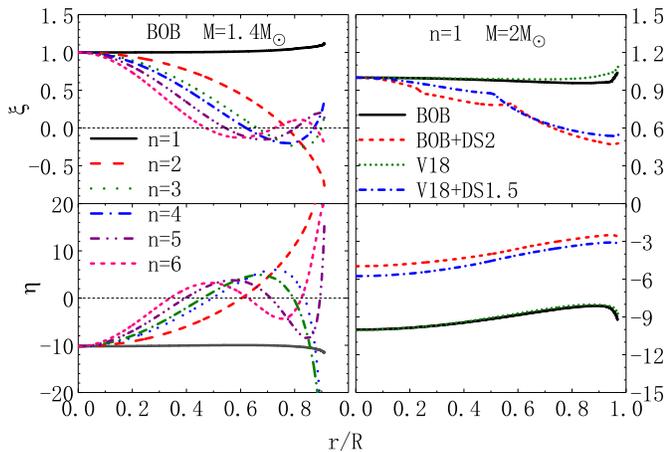}}
\vskip-2mm
\caption{
The radial displacement perturbation $\xi=\Delta r/r$ (upper panels)
and the radial pressure perturbation $\eta=\Delta p/p$ (lower panels)
of the first six eigenmodes of radial oscillations in $1.4\ms$ NSs
with BOB EOS (left panels)
and the fundamental eigenmode in $2.0\ms$ stars
with different EOSs (right panels).
}
\label{f:bob}
\end{figure}

\begin{figure}[t]
\centerline{\includegraphics[width=0.48\textwidth]{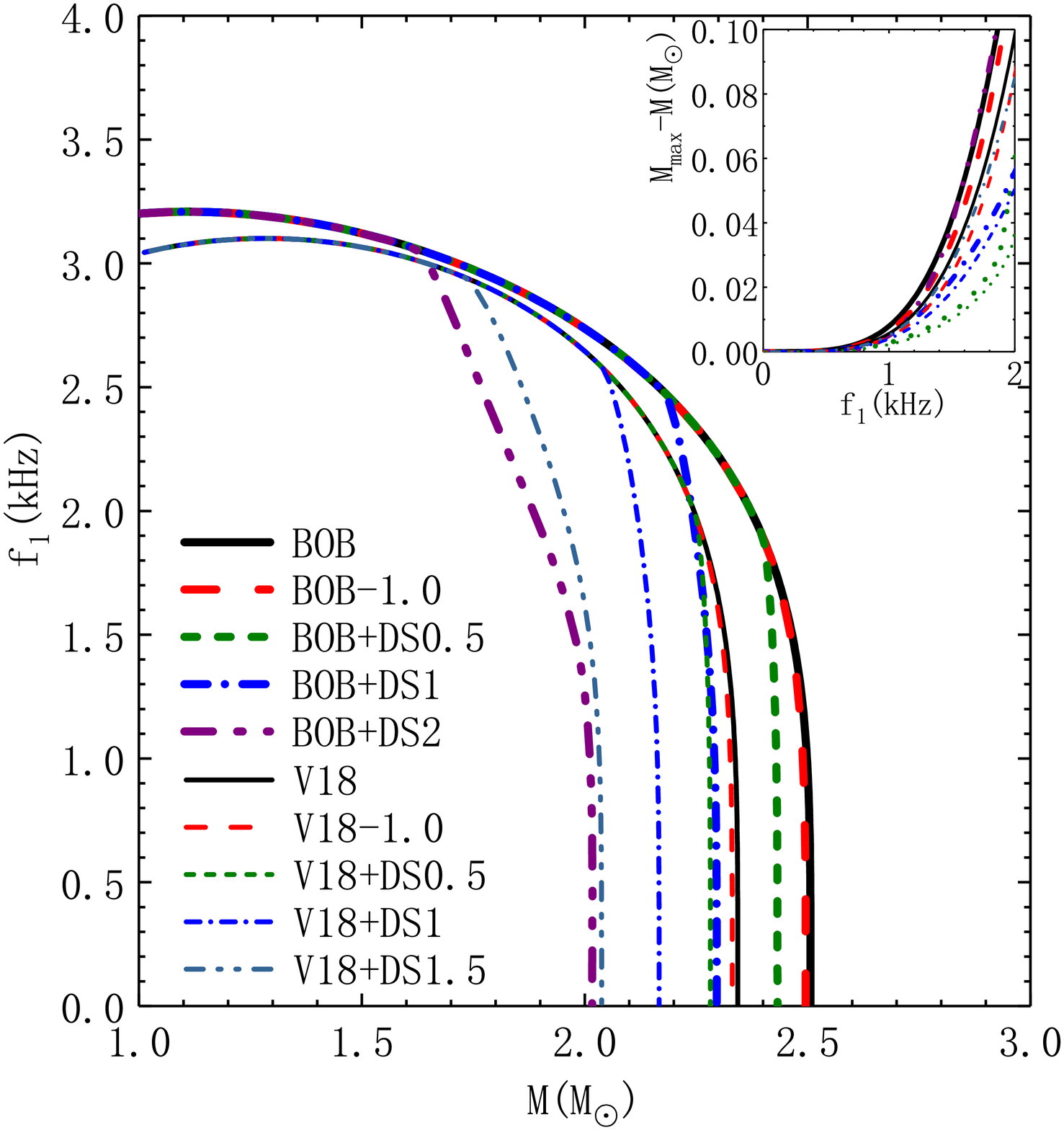}}
\vskip-4mm
\caption{
Fundamental frequency $f_1$ vs mass $M$,
and (inset) $M_\text{max}-M$ vs $f_1$
for different EOSs.
The notation is as in Fig.~\ref{f:sound}.}
\label{f:f1}
\end{figure}

\subsection{Radial oscillations and stability}

We now investigate the radial oscillations of NSs.
In the left panels of Fig.~\ref{f:bob} we show the first six eigenmodes
in a $1.4\ms$ NS, obtained with the BOB EOS.
There is no zero point in the first (fundamental) eigenmode.
As the order increases,
the number of zero points increases,
as is expected in a Sturm-Liouville boundary problem.
It can be seen that in the fundamental mode $n=1$
the entire star oscillates nearly uniformly,
whereas in the higher modes only the inner parts tend to be affected.
In the right panels of Fig.~\ref{f:bob} we show the fundamental eigenmodes
in a $2.0\ms$ NS, obtained with the various EOSs.
For pure NSs, the fundamental eigenmodes are almost the same
as in a $1.4\ms$ NS,
i.e., uniform oscillation.
In HSs there is a kink in $\xi$ due to the emergence of the mixed phase
and the $\xi$ decreases much faster towards the surface,
i.e., the oscillation is stronger in the QM phase.
The kink in $\xi$ is due to the large discontinuity of the adiabatic index,
which appears explicitly in Eq.~(\ref{e:xi}).
This causes a related kink in $d\eta/dr$, Eq.~(\ref{e:dp}).
There should also be a kink with the emergence of muons,
but quantitatively too small to be seen.

\begin{table}[t]
\caption{
The radial oscillation frequencies $f_n\,$[kHz]
of $M/\ms=1.4,2.0$ NSs
using different EOSs.}
\begin{ruledtabular}
\begin{tabular}{c|rr|rrrr}
 \multirow{2}{*}{$n$} &
 \multicolumn{2}{c|}{$1.4\ms$} &
 \multicolumn{4}{c}{$2.0\ms$} \\
   & BOB & V18 & BOB & BOB+DS2 & V18 & V18+DS1.5 \\
\hline
 1 &  3.16 &  3.09 &  2.73 &  1.25 &  2.64 &  1.59 \\
 2 &  6.74 &  6.65 &  6.80 &  5.90 &  6.85 &  6.11 \\
 3 &  8.81 &  9.25 & 10.08 &  8.57 & 10.05 &  9.16 \\
 4 &  9.69 & 10.15 & 12.24 & 11.44 & 12.93 & 11.78 \\
 5 & 11.59 & 11.90 & 13.45 & 13.39 & 14.72 & 14.36 \\
 6 & 13.83 & 14.20 & 15.08 & 14.51 & 16.04 & 15.64 \\
\end{tabular}
\end{ruledtabular}
\label{t:om}
\end{table}

The corresponding radial oscillation eigenfrequencies
are listed in Table~\ref{t:om}.
For a $1.4\ms$ NS we obtain the frequencies of the $n=1$ fundamental mode
$f_1=3.16/3.09\;$kHz with BOB/V18 EOS,
which are quite close in the two models.
Comparing with the previous literature \cite{Sagun2020},
our results of $f_1$ are quite similar,
but frequencies of higher modes $f_n$ are relatively smaller.
According to the features of the eigenmodes in Fig.~\ref{f:bob},
the fundamental frequency is determined by properties throughout the whole star,
whereas the higher frequencies depend mainly on the core properties.
The fundamental frequency is the easiest to be observed
in the next generation of GW detectors,
but we expect also the first few lower frequencies to be observable.
For a $2.0\ms$ NS $f_1$ depends much stronger on the EOS,
in particular, the $f_1$ of HSs are much smaller than those of pure NSs,
while $f_{n>1}$ increase more quickly.
This is compatible with the much lower speed of sound in HSs,
see Fig.~\ref{f:gamma}.

To investigate more the relation between radial oscillations and stability of NSs,
we show the dependence of the radial oscillation frequencies $f_1$
on the masses of NSs in Fig.~\ref{f:f1}.
One can see that $f_1$ varies slowly around 1.4 solar masses,
but decreases quickly close to the maximum mass.
It decreases to zero exactly at the numerical maximum mass on the $M-R$ curves,
when the star does not recover anymore from a small radial perturbation.
That is to say,
the commonly used stability condition $dM/d\rho_c>0$
\cite{Harrison:1965,shapiro}
is consistent with the analysis of the radial oscillations.
This is the case for pure NSs and for HSs.
Since $f_1$ decreases quickly close to the maximum mass,
one can quite exactly predict the maximum mass
when observing sufficiently small values of $f_1$.
For example,
as shown in the inset of Fig.~\ref{f:f1},
when observing a NS with $f_1=1.6\;$kHz and mass $M$,
one can predict that $\mmax<M+0.05\ms$.
(This is the situation for both HSs with $2\ms$ in Table~\ref{t:om},
for example).
For $f_1=1.0\;$kHz the constraint becomes even much stronger
and yields $\mmax<M+0.01\ms$.
A low oscillation frequency thus provides great added value to
a NS mass measurement,
which by itself only represents a lower limit on $\mmax$ of NSs.

\begin{figure}[t]
\centerline{\includegraphics[width=0.48\textwidth]{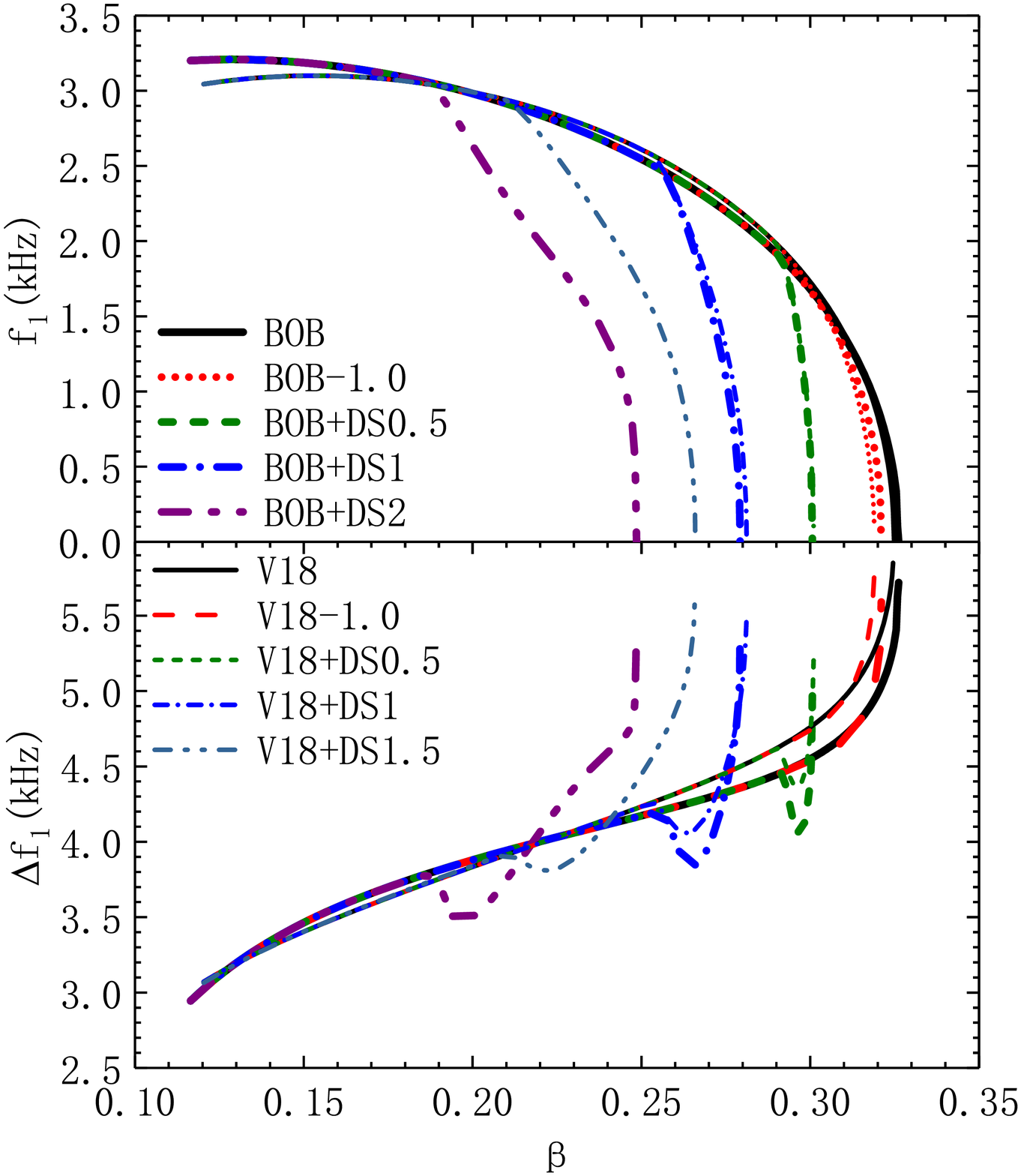}}
\vskip-4mm
\caption{
The dependence of $f_1$ (upper panel)
and $\Delta f_1=f_2-f_1$ (lower panel)
on the compactness parameter $\beta$
for different EOSs.
The notation is as in Fig.~\ref{f:sound}.}
\label{f:df1}
\end{figure}

It is also interesting to study the dependence of $f_1$
on the compactness parameter $\beta$,
shown in the top panel of Fig.~\ref{f:df1}.
We find that this dependence
is quite insensitive to the EOS of hadronic matter.
Even in the HSs,
the two curves with the same QM EOS combined with different hadronic EOSs
almost coincide.
In our approach,
HSs close to the mass limit
(indicated by small values of $f_1$)
have smaller masses and larger radii than pure NSs
(see Fig.~\ref{f:mr}),
and therefore their maximum compactness is smaller.
Thus a low value of $\beta$ ($\lesssim 0.3$)
together with a low value of $f_1$ ($\lesssim 1\;$kHz)
is an indication for the presence of QM inside the star.
This qualitative difference between HSs and pure NSs offers us
an important observational signal to disentangle them.

A similar feature applies to
the so-called large separations
$\Delta f_n \equiv f_{n+1}-f_n$,
which are widely used in astroseismology to learn about star properties
\cite{Sagun2020}.
We show the results of $\Delta f_1=f_2-f_1$
in the lower panel of Fig.~\ref{f:df1}.
Again one can find that the curves are insensitive to the EOS of hadronic matter,
and that in this case large values of $\Delta f_1\gtrsim 5\;$kHz
together with small values of $\beta\lesssim 0.3$
indicate substantial QM content.

Radial oscillations usually occur during strong transitions of NSs,
e.g., in newborn NSs after supernova explosions or the merger of NSs,
after strong starquakes and corresponding pulsar glitches.
The signals can be observed in the associated electromagnetic
and GW emissions.
In Ref.~\cite{Chirenti2019},
the authors point out that the radial oscillation can modulate
the SGRB from the hypermassive NSs formed after the merger of two NSs.
The possible observation of the radial oscillation frequencies in SGRB
will provide insight into the emission mechanism of the SGRB
and aid to understanding the EOS of NSs.
Radial oscillations can also couple with and amplify GWs
\cite{Passamonti2005,Passamonti2007}.

Although the frequency of the radial oscillation is too high
for the current GW detectors
(but may become sufficiently low for metastable hypermassive remnants),
it could be possibly observed with the improvement of the detectors,
such as the Advanced LIGO \cite{Regimbau2016},
and the third-generation ground-based GW detectors
such as the Einstein Telescope \cite{Punturo2010}
and Cosmic Explore \cite{Evans2016}.

\bigskip
\section{Conclusions}
\label{s:end}

In this work we investigated the radial oscillations and stability of NSs,
including both pure NSs and HSs.
The EOS of nuclear matter is based on the BHF theory,
limiting the speed of sound to the speed of light at high density.
Alternatively, we considered the phase transition to QM,
combining the nuclear EOS with a DSM EOS for QM
via a Gibbs phase transition.
With these EOSs, we solved the TOV equation for the equilibrium structure
and the equations for the radial oscillations of NSs.
For a $1.4\ms$ NS we obtain radii around $12\,$km
and $f_1\sim 3\,$kHz for the fundamental radial oscillation.
As the masses increase,
$f_1$ decreases to zero at the maximum mass,
consistent with the stability criterion $dM/d\rho_c=0$.
Small values of $f_1$ can provide an accurate estimate of the maximum NS mass.
Also, small values of the compactness parameter $\beta$
together with small values of $f_1$ or large values of $\Delta f_1$
characterize HSs in our approach
and allow to disentangle them from pure NSs.

To investigate radial oscillations in newborn NSs
after supernova explosions or the merger of NSs,
the features of more realistic environments, e.g.,
temperature, rotation, and magnetic field,
should also be included \cite{Panda2016}.
We leave these studies to future work,
together with those employing other EOSs.

\begin{acknowledgments}

We acknowledge financial support from the National
Natural Science Foundation of China (Grant Nos. 11475149
and U1738130).

\end{acknowledgments}

\newcommand{\epja}{Euro. Phys. J. A\ }
\newcommand{\apjl}{Astrophys. J. Lett.\ }
\newcommand{\mnras}{Mon. Not. R. Astron. Soc.\ }
\bibliography{nsosc}

\end{document}